\documentclass[preprint,seceq]{jpsj2}

\title{
On the origin of multiple ordered phases in PrFe$_4$P$_{12}$}

\author{Annam\'{a}ria \textsc{Kiss}\thanks{E-mail address: amk@cmpt.phys.tohoku.ac.jp} 
 and Yoshio \textsc{Kuramoto}\thanks{E-mail address: kuramoto@cmpt.phys.tohoku.ac.jp} 
}

\inst{%
Department of Physics, Tohoku University, Sendai 980-8578}

\recdate{\today}

\abst{
The nature of 
multiple electronic orders in skutterudite PrFe$_4$P$_{12}$ is 
discussed on the basis of a model with antiferro-quadrupole (AFQ) interaction of $\Gamma_3$ symmetry.
The high-field phase can be reproduced qualitatively provided 
(i) ferro-type interactions are introduced between the dipoles as well as between the octupoles 
of localized $f$-electrons, and (ii) separation is vanishingly small  between the $\Gamma_1$--$\Gamma_4^{(1)}$ crystalline electric field (CEF) levels.
The high-field phase can have either the same ordering vector ${\bf q}=(1,0,0)$ as in the low-field phase, or a different one ${\bf q}=0$ depending on the parameters. 
In the latter case, distortion of the crystal perpendicular to the $(111)$ axis is predicted.  The corresponding anomaly in elastic constants should also appear.
The electrical resistivity is calculated with account of scattering within the CEF quasi-quartet.
It is found that the resistivity as a function of the direction of magnetic field 
shows a sharp maximum around the $(111)$ axis at low temperatures because of the level crossing.}

\kword{%
skutterudite, quadrupole order, octupole, crystalline electric field, PrFe$_4$P$_{12}$
}

\begin{document}
\sloppy
\maketitle

\section{Introduction}

Rare earth filled skutterudites are intensively studied in the
last few years because of their variegated and complex behavior.
In this paper we focus among others on 
PrFe$_{4}$P$_{12}$ which 
exhibits a phase transition at 6.5K. 
A fascinating new experimental result on 
PrFe$_{4}$P$_{12}$ is the discovery of a high-field ordered phase sharply located around the 
$(111)$ direction of the magnetic field\cite{tayama2}. 
It is natural to expect that this new phase is related to the strange angle dependence of the
electrical resistivity which shows a sharp peak around the $(111)$ field
direction \cite{kuramochi}. 
The purpose of this paper is to provide a consistent account of the phase diagram of PrFe$_4$P$_{12}$ including both low- and high-field orders.  
On the basis of theoretical analysis, we predict  possible lattice distortion, 
which leads to a symmetry lower than trigonal along $(111)$, 
 and elastic anomaly on entering the high field phase.

Pr ions in the PrFe$_{4}$P$_{12}$ filled skutterudite have 4$f^2$
configuration with total angular momentum $J=4$. 
Under the point-group symmetry $T_h$,  the ninefold degenerate $J=4$ level is split into CEF levels.
The nature of the low energy CEF levels is essential
to the physical properties and electronic orderings of the
system. 
The CEF splittings in related systems PrRu$_4$P$_{12}$ and PrOs$_4$Sb$_{12}$ are clearly known by neutron diffraction experiments\cite{iwasa10,kuwahara2}. 
However, the sequence of CEF levels in PrFe$_{4}$P$_{12}$
has not yet been established experimentally.
Interpretation of neutron scattering spectra of PrFe$_4$P$_{12}$ remains ambiguous: polycrystalline samples do not show clear peaks and the peaks appearing in the single crystalline samples become clear only below the transition temperature\cite{iwasa11}.   
The first attempt on determining the CEF scheme was the analysis of the magnetization anisotropy in the disordered phase\cite{aoki}. 
It was found that the anisotropy can be reproduced either by schemes with $\Gamma_1$--$\Gamma_4$, $\Gamma_1$--$\Gamma_5$ or $\Gamma_3$--$\Gamma_4$ low-lying levels within the point group $O_h$.
Measurements of elastic constant  found 
softening of the modes $C_{11}$ and $C_{11}-C_{12}$ towards low temperatures\cite{nakanishi}. 
The results were interpreted in terms of the $\Gamma_3$ doublet ground state.


We find that the $\Gamma_1$--$\Gamma_4^{(1)}$ 
 scheme in magnetic fields gives rise to a level crossing only for field direction $(111)$.  
Our result generalizes the observation by Tayama et al.\cite{tayama2}, who worked with the model with the point group $O_h$. 
We thus consider that the relevant CEF scheme in PrFe$_{4}$P$_{12}$ is
the $\Gamma_1$--$\Gamma_4^{(1)}$ singlet-triplet scheme,
which is in line with ref.\citen{otsuki}. 
This level crossing is the key aspect of the high-field phase and will be fully discussed in Section~\ref{ssec:mult}.
If one takes the doublet ground state, on the other hand, the energy levels in magnetic field are rather different. With the doublet ground state, it is difficult to explain the high-field phase in terms of the level crossing.  
Analyzing the quadrupole pattern in the ordered phase at low temperatures, we shall clarify that a two-sublattice AFQ order of the $\Gamma_3$ quadrupoles within the $\Gamma_1$--$\Gamma_4^{(1)}$ subspace has a unique feature that does not share with a quadrupolar order in other CEF states such as the $\Gamma_3$ doublet or the 
 $\Gamma_8$ quartet. 


This paper is organized as follows. In Section~\ref{sec:2} single-site properties of our model is discussed.
In Section~\ref{ssec:mult} 
we examine the $\Gamma_3$ quadrupole interaction model for the field direction $(111)$ in detail. 
The effects of ferro-type dipolar and octupolar interactions on the phase boundary are analyzed.
We also derive the Curie-Weiss temperature.
In Section~\ref{sec:res} the electrical resistivity is calculated using the quasi-quartet 
crystal field states of $\Gamma_1$--$\Gamma_4^{(1)}$ 
as a function of ${\bf H}\|(111)$, 
and the angle dependence around the $(111)$ direction is derived.
In Section~\ref{sec:g3o} some basic features of the $\Gamma_3$ quadrupolar ordered phase in the triplet is discussed.
Section~\ref{sec:sum} is the summary and discussion of our results.

\section{Singlet-Triplet CEF states and their splitting in magnetic field}\label{sec:2}

\subsection{Crystal-field states and $\Gamma_3$ quadrupoles}
We start with the single-site Hamiltonian:
\begin{equation}
{\cal H}_{\rm ss}={\cal H}_{\rm CEF}-g\mu_{\rm B} {\bf H}\cdot {\bf J} ,
\end{equation}
where the second term represents the Zeeman energy with the angular momentum {\bf J}, and 
the first term is the CEF potential under
the tetrahedral crystal field.  The CEF term can be expressed as \cite{takegahara}
\begin{eqnarray}
{\cal H}_{\rm
CEF}=W\left(xO_4+(1-x)O_6^{c}+yO_6^{t}\right)\,,\label{eq:th}
\end{eqnarray}
where we have used the normalization convention $O_4^4=(J_{+}^4+J_{-}^4)/120$ with $J_{\pm}=J_{x}\pm iJ_{y}$ in $O_{4}\equiv O_{4}^0+5O_{4}^4$.

The term $yO_6^{t}$ has no effect on the cubic $\Gamma_1$ and
$\Gamma_3$ levels, but mixes the cubic $\Gamma_4$ and $\Gamma_5$
triplet states. Then the new triplets $\Gamma_4^{(1)}$ and $\Gamma_4^{(2)}$ arise.
The $\Gamma_4^{(1)}$ triplet can be written in terms of $O_h$ triplets
$\Gamma_4^{(0)}$ and $\Gamma_5^{(0)}$ as
\begin{eqnarray}
|\Gamma_4^{(1)},m\rangle=d|\Gamma_4^{(0)},m\rangle+\sqrt{1-d^2}|\Gamma_5^{(0)},m\rangle\,,
\end{eqnarray}
where
\begin{eqnarray}
d=\left[\frac{1}{2}\left(1+\frac{3+2x}{\sqrt{(3+2x)^2+1008y^2}}\right)\right]^{1/2},\label{eq:s}
\end{eqnarray}
with $x>0$ being assumed. The $O_h$ triplets are
\begin{eqnarray}
&& |\Gamma_4^{(0)},0 \rangle=\frac{1}{2}(|+4\rangle-|-4\rangle)\nonumber\\
&& |\Gamma_4^{(0)},\pm \rangle=\pm \sqrt{\frac{1}{8}}|\pm 3\rangle \pm \sqrt{\frac{7}{8}}|\mp 1\rangle
\end{eqnarray}
and
\begin{eqnarray}
&& |\Gamma_5^{(0)},0 \rangle=\frac{1}{2}(|+2\rangle-|-2\rangle)\nonumber\\
&& |\Gamma_5^{(0)},\pm \rangle=\mp \sqrt{\frac{7}{8}}|\mp 3\rangle \pm \sqrt{\frac{1}{8}}|\pm 1\rangle.
\end{eqnarray}
We further introduce the notations for future convenience
\begin{eqnarray}
|\Gamma_4^{(1)},0\rangle \equiv |0\rangle, \hspace*{5mm}|\Gamma_4^{(1)},\pm \rangle \equiv |\pm\rangle.
\end{eqnarray}
The energies of the states of the pseudo-quartet are given by 
\begin{eqnarray}
E(\Gamma_4^{(1)})&=&
2W\left[x-4+2\sqrt{(3+2x)^2+1008y^2}\right]\,,\nonumber\\
E(\Gamma_1)&=& W\left[108x-80\right]\,.\label{eq:eng1g4}
\end{eqnarray}
We note that 
the parameter $y$ 
is indispensable for getting the $\Gamma_1$ and $\Gamma_4^{(1)}$ states closely located to each other without the proximity of the $\Gamma_3$ doublet state.

Now, let us consider the $\Gamma_3$ quadrupole moments under the tetrahedral symmetry $T_h$.
As an alternative to 
the pseudo-spin representation\cite{shiina2},
we directly compare matrix elements of
the quadrupolar operators $O_2^0=[3J_z^2-J(J+1)]/\sqrt{3}$ and 
$O_2^2=J_x^2-J_y^2$ with respect to 
the tetrahedral $\Gamma_4^{(1)}$ triplet states 
$\{
|\alpha_1\rangle, |\alpha_2\rangle, |\alpha_3\rangle\}=\{
|0\rangle,|+\rangle,|-\rangle\}$, and 
the  cubic $\Gamma_4$ triplet states $|a_{i}\rangle$ ($i=1,2,3)$.
Then we define the $3\times 3$ matrices
${\cal O}^\mu_2$ and ${\cal C}^\mu_2$
with $\mu=0,2$ by 
\begin{eqnarray}
({\cal O}^\mu_2)_{ik} = \langle \alpha_{i}|O^\mu_2|\alpha_k \rangle, \ \ 
({\cal C}^\mu_2)_{ik} = \langle a_{i}|O^\mu_2|a_k \rangle. 
\end{eqnarray}
We obtain the matrix equation
\begin{eqnarray}
{{\cal O}_2^0} &=& a_Q {\cal C}_{2}^0 + b_Q {\cal
C}_{2}^2\,,\nonumber\\
{{\cal O}_2^2} &=& a_Q {\cal C}_{2}^2 - b_Q {\cal
C}_{2}^0\,,\label{eq:g3k2}
\end{eqnarray}
where $a_Q=(9d^2-2)/7$ and $b_Q=-d\sqrt{3(1-d^2)/7}$. 
The limits $d=1$ and 0 mean the cubic
triplets $\Gamma_4$ and $\Gamma_5$, respectively.
Namely we recover 
${\cal O}_2^0={\cal C}_{2}^0$ and ${\cal O}_2^2={\cal C}_{2}^2$ 
from (\ref{eq:g3k2})  because of $a_Q=1$ and
$b_Q=0$. 
The inverse transformation of equations (\ref{eq:g3k2}) gives 
\begin{eqnarray}
{\cal C}_{2}^0 &=&
\frac{a_Q}{(a_Q^2+b_Q^2)} \ {\cal O}_{2}^0 -
\frac{b_Q}{(a_Q^2+b_Q^2)} {\cal O}_{2}^2\,, \nonumber\\
{\cal C}_{2}^2 &=& 
\frac{a_Q}{(a_Q^2+b_Q^2)} {\cal O}_{2}^2 +
\frac{b_Q}{(a_Q^2+b_Q^2)} {\cal O}_{2}^0\,. \label{eq:g3k3}
\end{eqnarray}
We remark that 
${\cal O}_{2}^\mu$ with $\mu=0,2$
can be diagonalized simultaneously by the basis set $|\pm\rangle, |0\rangle$.
The eigenvalues can be obtained from the eigenvalues 
$-14/\sqrt{3} (1,1,-2)$
of ${\cal C}_{2}^0$, 
together with eigenvalues $\pm 14$, $0$
of ${\cal C}_{2}^2$ by proper combination.
Using equations (\ref{eq:g3k3}) we 
obtain 
\begin{eqnarray}
({\cal O}_{2}^0)^2+({\cal O}_{2}^2)^2 =
(a_Q^2+b_Q^2)[
({\cal C}_{2}^0)^2+({\cal C}_{2}^2)^2] = \frac{784}{3}(a_Q^2+b_Q^2),
\end{eqnarray}
which is a scalar matrix in consistency with the fact  that $(O_2^0)^2+(O_2^2)^2$
is a Casimir operator.
The factor $a_Q^2+b_Q^2 = (4-15d^2+60d^4)/49$ reduces to unity with $d=1$, and to 4/49 with $d=0$.  Namely the $\Gamma_5^{(0)}$ triplet has a smaller
quadrupole moment as compared with the $\Gamma_4^{(0)}$ triplet.

\subsection{Constraints on the singlet-triplet CEF splitting}

Analyzing the energy spectra of the Hamiltonian 
${\cal H}_{\rm ss}={\cal H}_{\rm CEF}-g\mu_{\rm B} {\bf H}\cdot {\bf J}$ 
for different crystal
field parameters, 
we can easily find that only the $\Gamma_1$--$\Gamma_4^{(1)}$ low-lying scheme gives a crossing
point uniquely for the $(111)$ direction of the magnetic field.
The mixing parameter 
$d$ has to be larger than $d_{cr}=\sqrt{15}/6\approx 0.645$,
because for $d<d_{cr}$ a level crossing occurs also for ${\bf H}\|(001)$.
This conclusion comes from the comparison of the inter- and intra-level matrix elements of $J_z$ 
within the $\Gamma_1$--$\Gamma_4^{(1)}$ quasi-quartet.
These are $\pm 2\sqrt{15}d/3$ and $\pm (5/2-2d^2)$, respectively. 

Furthermore, 
the existence of the high-field phase imposes a strong constraint on the singlet-triplet CEF splitting.
With the singlet CEF ground state, 
the quadrupolar order must be an interaction induced order. We need a finite coupling constant  
between the quadrupoles for the phase transition even at zero temperature.  
In the case of a large energy separation $\Delta$, 
on the other hand, the transition changes to first-order from second-order for $\Delta>\Delta_{tcr}$, where $\Delta_{tcr}$ means the tricritical value of the energy separation. Loosely speaking, in a mean-field theory we expect that $\Delta_{tcr}$ has the same order of magnitude as the transition temperature $T_Q=6.5$K. Experimentally the phase transition at $T_Q$ is second-order, hence we obtain the upper bound of the CEF splitting of about 7K within our model.
On the other hand, the high-field phase occurs with infinitesimal interaction because of the level crossing.  If the interaction parameters responsible for the low field AFQ order works also in the high-field phase, the observed phase diagram suggests very small splitting of the quasi-quartet.
Otherwise, the large interaction necessary for the induced AFQ order makes the high-field phase much wider than actually is.
The critical field $H_{cr}$ becomes also too large. 

\begin{figure}[ht]
\centering
\includegraphics[totalheight=9cm,angle=270]{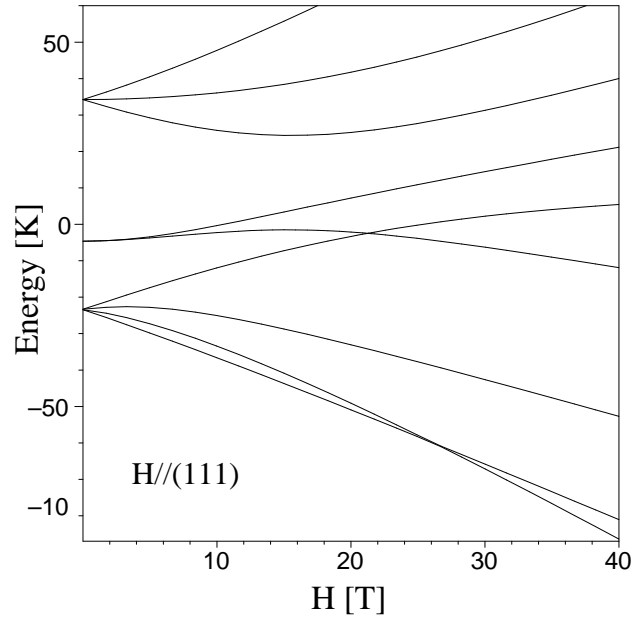}
\caption{Magnetic field evolution of the CEF states with $W=-0.9$K, $x=0.98$ and $y=-0.197$, which corresponds to $d=0.9$. The ground state at $H=0$ is the $\Gamma_1$--$\Gamma_4^{(1)}$ quasi-quartet, 
and the first excited state is the $\Gamma_3$ doublet with $18.8$K higher.}\label{fig:3}
\end{figure}

Therefore in the following we assume a singlet-triplet energy gap vanishingly small. This assumption $E(\Gamma_4^{(1)})-E(\Gamma_1)=0 $ 
defines a relation between the 
tetrahedral crystal field parameters $x$ and $y$ 
as given by eq.(\ref{eq:eng1g4}). 
Together with the constraint on $d$ discussed earlier, 
we choose $d=0.9$ which gives $x=0.98$ and $y=-0.197$, and
we use $W=-0.9$K. 
Figure~\ref{fig:3} shows evolution of the crystal field levels as the magnetic field increases with these parameters.
Increasing the strength of the magnetic field, a level crossing occurs in the ground state. We note that the behavior remains similar in the presence of small positive $\Delta$. 

\begin{figure}
\centering
\includegraphics[totalheight=7cm,angle=270]{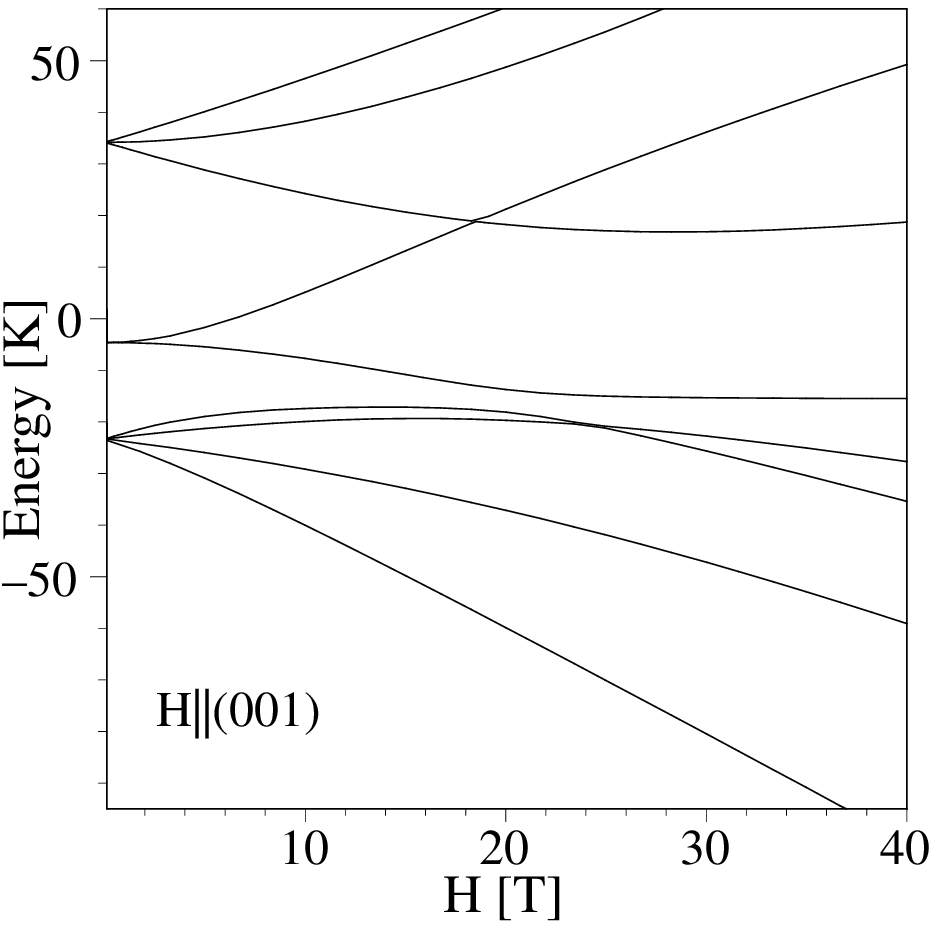}
\includegraphics[totalheight=7cm,angle=270]{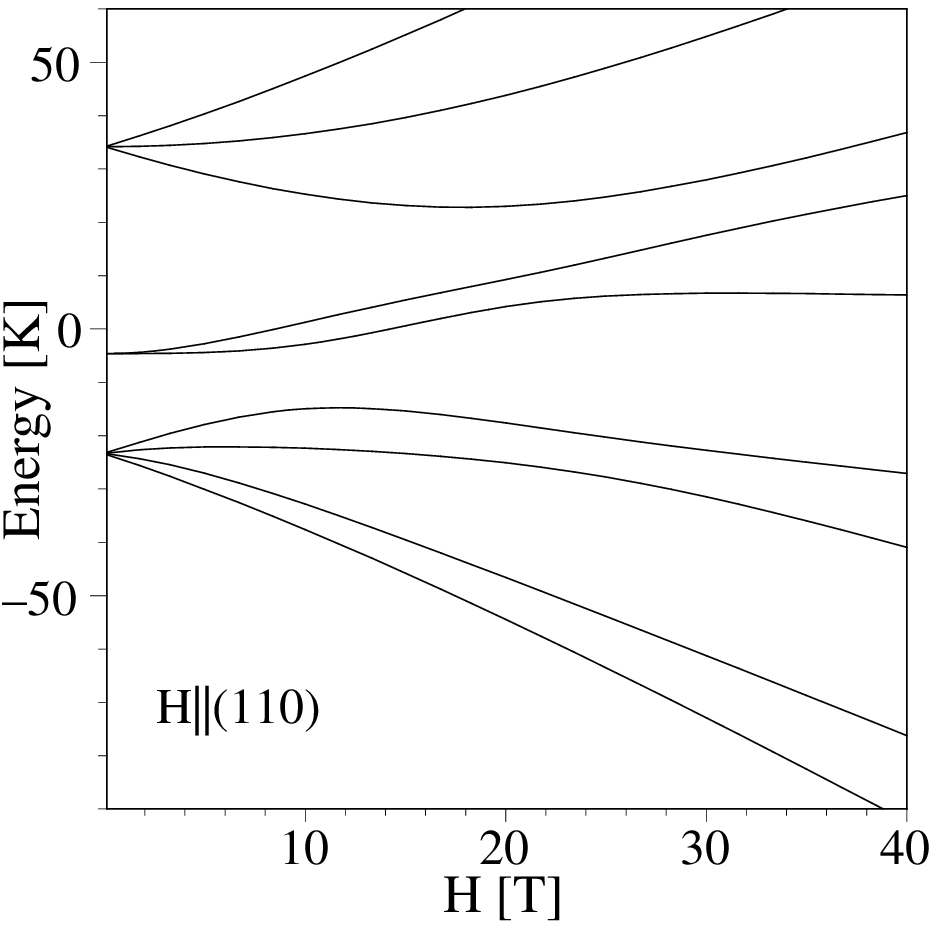}
\caption{Magnetic field evolution of the CEF states for field directions $(001)$ ({\sl left}) and $(110)$ ({\sl right}) with parameters $W=-0.9$K, $x=0.98$ and $y=-0.197$, which corresponds to $d=0.9$.}\label{fig:cef}
\end{figure}

We also derived the field evolution of the CEF states for field directions $(001)$ and $(110)$, which are shown in Fig.~\ref{fig:cef}. 
We can observe that no level crossing occurs in the ground state for these field directions. Therefore, the effect of application the magnetic field in these directions is expected to be the suppression of the antiferro-quadrupolar ordered phase.

%
%

\section{Phase diagram in the multipolar interaction model}\label{ssec:mult}
\subsection{Instability of the paramagnetic phase }

We shall work with the minimal model for intersite interactions which can reproduce the two ordered phases at least qualitatively.   
We include such octupoles that reduce to $\Gamma_{5u}$ in the cubic symmetry. These are given by 
$T^{\beta}_{x}=(\overline{J_xJ_y^2}-\overline{J_z^2J_x})/6$, $T^{\beta}_{y}=(\overline{J_yJ_z^2}-\overline{J_x^2J_y})/6$ and $T^{\beta}_{z}=(\overline{J_zJ_x^2}-\overline{J_y^2J_z})/6$, where the bar means symmetrization.
Then we consider the following model with the nearest-neighbor intersite interaction:
\begin{equation}
{\cal H}_{\rm int} 
=
\frac{1}{2}\sum_{\langle
i,j\rangle}\left[ \lambda_{\Gamma_3}
(O_{2,i}^0\cdot O_{2,j}^0+
O_{2,i}^2\cdot O_{2,j}^2)+
\lambda_{oct} {\bf T}^{\beta}_i 
\cdot {\bf T}^{\beta}_j+\lambda_{dip}{\bf J}_i\cdot {\bf J}_j\right],\label{eq:int3}
\end{equation}
with  $\lambda_{\Gamma_3}>0$ and $\lambda_{oct}<0,\ \lambda_{dip}<0$. Namely we take antiferro-type quadrupolar and ferro-type dipolar and octupolar interactions. 
The dipole and $\Gamma_3$ quadrupolar interactions are consistent with the invariant form allowed by symmetry of the bcc lattice \cite{sakai1}.
On the other hand, for simplicity we have neglected the mixing term  
${\bf J}\cdot{\bf T}^\beta=J_xT^{\beta}_{x}+J_yT^{\beta}_{y}+J_zT^{\beta}_{z}$
between the dipoles and $T^{\beta}$ octupoles, 
although the mixing is allowed by the $T_h$ tetrahedral symmetry. 
%
In the following we constrain the magnitudes of parameters such as
the coupling constants and the CEF splitting.

We take the model given by ${\cal H}_{\rm ss} +{\cal H}_{\rm int}$,
and  obtain the phase boundary in the plane of magnetic field and temperature. 
The nine crystal field levels labeled by $| l\rangle$ 
are derived for different values of magnetic field by solving 
${\cal H}_{\rm ss} | l\rangle =E_{l}| l\rangle$. 
In deriving the phase boundary,  
we keep only the lowest six levels for simplicity of the numerical calculation. 
As seen from Fig.~\ref{fig:3}, the three higher levels neglected here repel with lower levels, and go up in energy.
Thus their neglect will not influence the low-temperature behavior significantly. 

From the six levels labelled by $|l\rangle$ with $l=1,...,6$, 
we have $35\ (=6\times 6-1)$ independent pairs $k,l$ describing
multipole operators at each site. 
We have 15 symmetric combinations of the pairs, and
arrange them as increasing order of $10k+l$ with $k<l$.
Then we define $X^{\alpha}=| k \rangle \langle l |+| l \rangle \langle k |$ 
with $\alpha=1,...,15$ according to the ordering of the pair.
Similarly we define 15 antisymmetric combinations
$X^{\beta}=i\cdot(| k \rangle \langle l |-| l \rangle \langle k |)$ with $\beta =16,...,30$. 
The remaining five operators are diagonal in $k$ and $l$.
The general form of the multipolar
interaction is given by
\begin{eqnarray}
{\cal H}_{\rm int}=-\frac{1}{2}\sum_{\langle
i,j\rangle}\sum_{\alpha,\beta}V^{\alpha\beta}_{ij}X^{\alpha}_{i}X^{\beta}_{j}=-\frac{1}{2}\sum_{\langle
i,j\rangle} {\bf X}^{T}_{i}\cdot
{\hat{V}}_{ij}\cdot {\bf X}_{j}\,,
\end{eqnarray}
where ${\bf X}=[X^{1},X^{2},..,X^{35}]$, and $i,j$ are site indices. 
The 
susceptibility matrix can be expressed as
\begin{eqnarray}
\hat{\chi}=\left(1- \hat{\chi}_0\cdot
\hat{V}\right)^{-1}\cdot
\hat{\chi}_0\,,\label{eq:nisus1}
\end{eqnarray}
where $\hat{\chi}_0$ is the 
single-site susceptibility matrix whose elements are given by 
\begin{eqnarray}
(\chi_0)^{\alpha\beta}=\sum_{k,l}\frac{\rho(E_{k})-\rho(E_{l})}
{E_{l}-E_{k}}\langle k |X^\alpha|l \rangle \langle l |X^\beta|k\rangle\,,\label{eq:nis}
\end{eqnarray}
where $\rho(E_{k})={\rm exp}(-\beta E_{k})/\sum_{k}{\rm exp}(-\beta E_{k})$.
In the right-hand side of eq.(\ref{eq:nis}), we only need terms with $k\neq l$,
since the interaction Hamiltonian (\ref{eq:int3}) does not have
the diagonal part of operators $X^{\alpha}$. 
The calculation of 
$(\chi_0)^{\alpha\beta}$ can easily be performed with use of the magnetic eigenstates, which are nondegenerate except at the level crossing point.
It is important here 
to include renormalization of the external magnetic field $h =g\mu_B H$.
Namely we consider the effective magnetic field given by
$h_{\rm eff}=h-z\lambda_{dip}\langle J_{111}\rangle$
where $\langle J_{111}\rangle$ depends on the effective field $h_{\rm eff}$ and temperature $T$.  
The effective field should be determined self-consistently\cite{note2}.  
Actually we first take $h_{\rm eff}$ as a given field and 
derive the matrix $\hat{\chi}_0$ under $h_{\rm eff}$.
To each value of $h_{\rm eff}$, we determine $h$ by the relation $h=h_{\rm eff}+z\lambda_{dip}\langle J_{111}\rangle$.

After Fourier transformation we obtain the 
susceptibility matrix as a function of {\bf q}:
\begin{eqnarray}
\hat{\chi}({\bf q})=\left(1-\hat{\chi}_0\cdot
\hat{V}({\bf q})\right)^{-1}\cdot
\hat{\chi}_0\,,\label{eq:suscq2}
\end{eqnarray}
where 
$V^{\alpha\beta}({\bf q})$ 
is the Fourier transform of 
$V^{\alpha\beta}_{ij}$. 
The interaction matrix $V^{\alpha\beta}({\bf q})$ can be 
classified in a form 
\begin{eqnarray}
V^{\alpha\beta}({\bf q})=V^{\alpha\beta}_{\Gamma_3}({\bf q})+V^{\alpha\beta}_{oct}({\bf q})+V^{\alpha\beta}_{dip}({\bf q}). 
\end{eqnarray}
A second-order phase transition is characterized by the divergence of the 
 susceptibility matrix. This is equivalent with condition
\begin{eqnarray}
{\rm det}\left(1-\hat{\chi}_0\cdot
\hat{V}({\bf q})\right)=0\,,\label{eq:det}
\end{eqnarray}
where $1$ means the 35-dimensional identity matrix.
The solution of eq.~(\ref{eq:det}) gives the instability condition of the paramagnetic phase.  Note that the phase boundary may not coincide with the instability line if there is a first-order transition.
In this paper, we do not go into detailed inspection about the order of the transitions. 


\subsection{Ordering vector and instability lines in magnetic field}

Under the reduced symmetry with finite magnetic field along the (111) direction,
the original $T_h$ system allows 
only two different order parameter symmetries: 
the $C_3(\Gamma_1)$ singlet and the $C_3(\Gamma_3)$ doublet representations \cite{shiina2}. 
Within the two-dimensional local Hilbert space, which is formed by the closest crystal field states near the level crossing point, 
the order parameters with  $C_3(\Gamma_3)$  symmetry have nonzero inter-level matrix elements. 
%
If the high-field phase 
has the same ordering vector as the low-field AFQ phase,  
both phases can be connected smoothly. 
Alternatively, the high-field phase can have a different ordering vector ${\bf q}$ without the smooth connection.

\begin{figure}
\centering
\includegraphics[totalheight=7cm,angle=270]{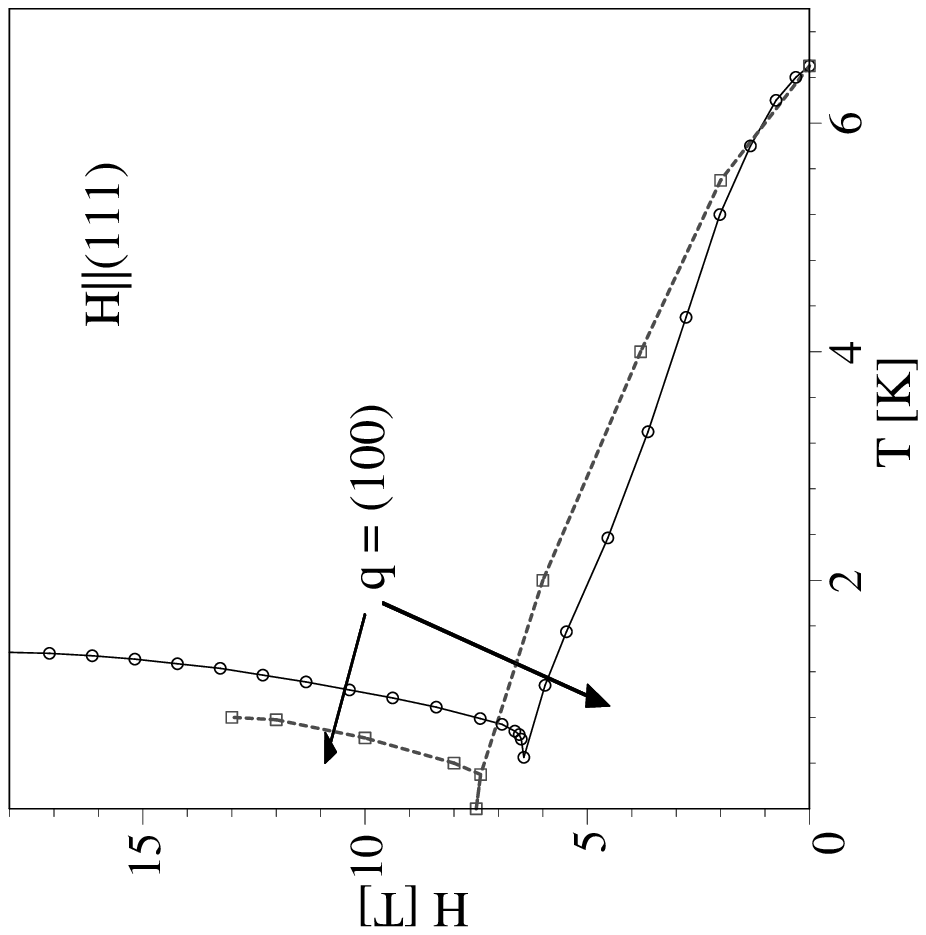}
\includegraphics[totalheight=7cm,angle=270]{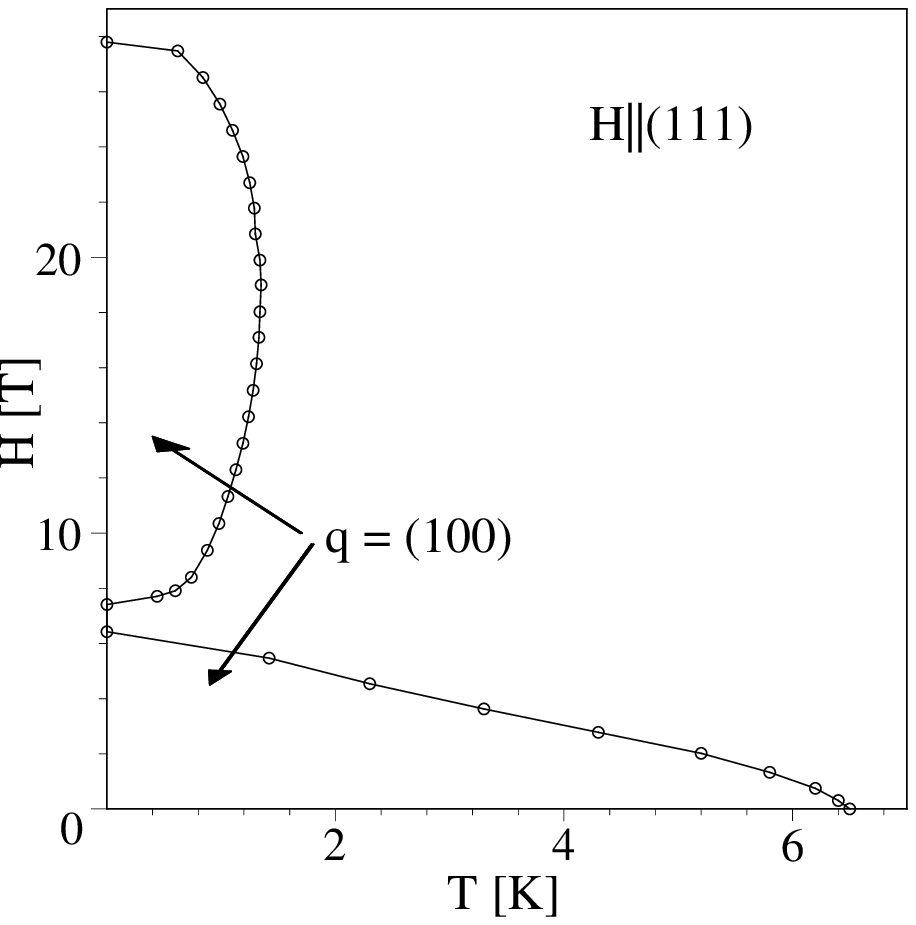}
\caption{Temperature--magnetic field phase diagram with interaction parameters 
$\lambda_{\Gamma_3}=8.93$mK
and $\lambda_{dip}=-125$mK.
The octupolar interaction is taken to be $\lambda_{oct}=-8.625$mK ({\sl right}) and 
$\lambda_{oct}=-8.55$mK ({\sl left}). On the left panel the dashed line shows the experimental phase boundary.}\label{fig:1}
\end{figure}

We derive the instability toward the low-field and the high-field phases from high temperatures with several sets of interaction parameters.
We study both cases with the ordering vector ${\bf q}=(1,0,0)$, which leads to a two-sublattice order on the bcc. lattice, and ${\bf q}=0$.
The quadrupolar coupling constant $\lambda_{\Gamma_3}$ is determined by the zero field transition temperature $T_Q=6.5$K. In order to determine this quadrupolar coupling constant
within the lowest six levels, let us consider the mean-field form of the Hamiltonian ${\cal H}_{\rm ss}+{\cal H}_{\rm int}$ at $H=0$: 
\begin{eqnarray}
{\cal H}_{\rm MF}(\Gamma_1-\Gamma_4^{(1)}-\Gamma_3)
=\Delta
\sum_{n=1,2}{|\Gamma_3^{n}\rangle \langle
\Gamma_3^{n}|}+z\lambda_{\Gamma_3} \left( \langle O_2^0
\rangle_{A(B)} O_2^0+ \langle O_2^2
\rangle_{A(B)} O_2^2\right)\,,\label{eq:ham1}
\end{eqnarray}
where the quasi-quartet level is taken as the origin of energy, and the energy of the $\Gamma_3$ level is $\Delta$. 
We perform the Landau
expansion of the free energy. The quadrupolar phase transition, which is second order at $H=0$, is determined by zero coefficient $\alpha_q$ of the quadratic term. We obtain
\begin{eqnarray}
\alpha_q=\frac{1}{2}z\lambda_{\Gamma_3}-z^2\lambda_{\Gamma_3}^2\frac{4}{3}\frac{[{\rm
exp}(-\Delta/T)(16\Delta-140T)+147(a_Q^2+b_Q^2)\Delta
+140T]}{T\Delta(4+2{\rm exp}(-\Delta/T))}\,.\label{eq:alfa1}
\end{eqnarray}
With the 
parameters used in Fig.~\ref{fig:3} we obtain $\lambda_{\Gamma_3}=8.93$mK
from $\alpha_q=0$.

\begin{figure}
\centering
\includegraphics[totalheight=7cm,angle=270]{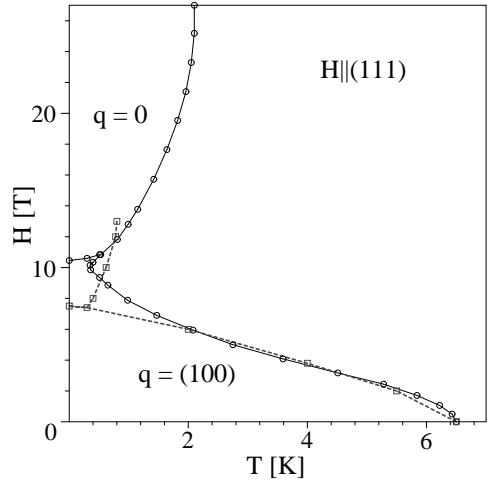}
\caption{Temperature--magnetic field phase diagram with 
$\lambda_{\Gamma_3}=8.93$mK, $\lambda_{dip}=-102.5$mK, and $\lambda_{oct}=-11.39$mK. Dashed line shows the observed phase boundary.}\label{fig:2}
\end{figure}

Two different solutions can be obtained depending on the 
magnitude of ferro-type coupling constants: 
(a) both low- and high-field phases have the same 
ordering vector ${\bf q}=(1,0,0)$; 
(b) the high-field phase has the ordering vector ${\bf q}=0$. 
In the case (a) we obtain either continuous phase boundary or two separated phases (see Fig.~\ref{fig:1}). 
For example, with $\lambda_{dip}=-125$mK and $\lambda_{oct}=-8.55$mK 
we obtain the continuous behavior.
While with the same $\lambda_{dip}$ but $\lambda_{oct}=-8.625$mK, 
two phases are separated.
%
%
An example for case (b) can be seen in Fig.~\ref{fig:2}, where 
the parameter are $\lambda_{dip}=-102.5$mK, $\lambda_{oct}=-11.39$mK.

The effects of the 
ferro-type interactions on the phase boundary depend on the multipolar matrix elements and the quadrupolar-dipolar-octupolar mixing due to the symmetry lowering as we increase the magnetic field. 
We found 
that the dipolar matrix elements are effective
in the reduction of the transition temperature in the low-field regime ($H<H_{\rm cross}$), while
the $T^{\beta}$ octupolar matrix elements have larger effects in the high-field regime ($H\sim H_{\rm cross}$). 
This latter behavior arises 
because the inter-level octupolar matrix elements between $\Gamma_4^{(1)}$--$\Gamma_{3}$ levels
become significant 
as we increase the magnetic field. 
In the case of larger ferro-octupolar coupling, 
the high-field phase thus takes ${\bf q}=0$ as the ordering vector.

\subsection{Magnetic susceptibility and Curie-Weiss temperature}\label{sec:weiss}

We examine whether the 
parameters used in the previous calculation 
are consistent with other observables.
For this purpose we 
discuss the temperature dependence of the magnetic susceptibility in the high-temperature phase.  
Comparsion of the Curie-Weiss temperature 
between theory and experiment will provide a test of our model.
The positive Weiss temperature found experimentally  
is the most straightforward evidence of the ferro-type interaction between the dipoles. 
We remark that 
a ferromagnetic phase transition has been found in Pr$_x$La$_{1-x}$Fe$_4$P$_{12}$ for $x=0.85$ \cite{aoki02}, which also indicates
the strong ferro-dipolar interaction. 
We have further introduced ferro-type interaction between the $T^{\beta}$ octupoles. In the tetrahedral symmetry, these octupoles have the same symmetry as the dipoles ($\Gamma_{4u}$), therefore they cannot be distinguished from the dipoles experimentally.

We start with the 
single-site susceptibility matrix.  
Instead of $X^\alpha$ in eq.(\ref{eq:nis}) as multipole operators, we can equivalently use another set of operators $Z^\mu$ with 
$Z^{1}=J_z$ and $Z^2 =T^{\beta}_z$. 
We obtain the $2\times 2$ matrix for the single-site susceptibility  as
\begin{eqnarray}
(\chi_0)^{\mu \nu}=\sum_{k}\rho(E_k)\left[-2\sum_{l\ne k}\frac{\langle
k|Z^{\mu}|l\rangle \langle l|
Z^{\nu}|k\rangle}{E_k-E_l}+\frac{1}{T} \langle k|
Z^{\mu}|k\rangle \langle k|
Z^{\nu}|k\rangle\right]. \label{eq:genchiojz}
\end{eqnarray}
The susceptibility in eq.(\ref{eq:genchiojz}) has the Curie term in 
contrast with eq.(\ref{eq:nis}).
%
Then the magnetic susceptibility $\chi_M$ including the intersite interactions
is given by
\begin{eqnarray}
\chi_M= \left[  
\hat{\chi}_0\left(1-\hat{\chi}_0\cdot
\hat{V}\right)^{-1}\right]_{11}\label{eq:nisus3}
\end{eqnarray}
where the matrix $\hat{V}$ has the form
\begin{eqnarray} \hat{V}= \left(
\begin{array}{cc}
           z\lambda_{dip} & 0\\
           0 & z\lambda_{oct} \end{array}
       \right)\,,
\label{eq:matrix}
\end{eqnarray}
with $z=8$. 
Because of the off-diagonal component $\chi_0^{12}$ and $\chi_0^{21}$, octupolar interaction influences the magnetic susceptibility.
The susceptibility given by (\ref{eq:nisus3}) diverges
at the Curie-Weiss temperature. 
This condition is equivalent with the divergence of the ${\bf q}=0$ component of the susceptibility (\ref{eq:suscq2}) in the absence of external magnetic field.

With the set of parameters used in the left panel of Fig.~\ref{fig:1}, namely
$\lambda_{dip}=-125$mK and  $\lambda_{oct}=-8.55$mK, 
we obtain $\Theta^{*}=4.41$K.
On the other hand, 
another set $\lambda_{dip}=-102.5$mK and $\lambda_{oct}=-11.39$mK, which is used in Fig.~\ref{fig:2}, gives $\Theta^{*}=4.12$K for the Curie-Weiss temperature.
These are to be compared with the experimental value $\Theta^{*} \sim 3.6$K\cite{aoki}. 

\section{Electrical resistivity in magnetic field}\label{sec:res}

In this Section we 
analyze the effect of the level crossing on the electrical resistivity.
We are especially interested in 
the sharp enhancement of resistivity around the $(111)$ field direction. 
We use a very simple model 
where conduction electrons are scattered by magnetic exchange and aspherical Coulomb interaction.
The change of population of the crystal field levels causes a temperature and magnetic field dependence in the resistivity. 
We follow the method presented in refs.\citen{maple,fisk},
and calculate the resistivity in the disordered phase.

Using the notation of ref.\citen{anderson}, the contribution to the resistivity from the CEF effects can be described as
\begin{eqnarray}
{\rho}_{\rm CEF}={\rho}_{0}[r{\rm Tr}(\hat{P}  \hat{Q}^{\rm M})+(1-r){\rm Tr}(\hat{P}\hat{Q}^{\rm A})]\,,
\end{eqnarray}
where $r$ means the ratio between the two scattering terms. The matrix $\hat{P}$ has the form
\begin{eqnarray}
P_{kl}=
\frac{\rho (E_k) \beta(E_k-E_l)}{1-{\rm e}^{-\beta (E_k-E_l)}}
=
\frac{\sqrt{\rho (E_k) \rho (E_l)}
\beta(E_k-E_l)}{2\cosh \frac 12 \beta(E_k-E_l)}
\end{eqnarray}
in the notation of eq.(\ref{eq:nis}).
%
The matrices $\hat{Q}^{\rm M}$ and $\hat{Q}^{\rm A}$ represent 
the magnetic exchange and aspherical Coulomb scattering, respectively.
The latter comes from the quadrupolar charge distribution of the Pr$^{3+}$ ions.
They are given by
\begin{eqnarray}
Q_{kl}^{\rm M} &=& |\langle k|J_x|l\rangle|^2+|\langle k|J_y|l\rangle|^2+|\langle k|J_z|l\rangle|^2\,,\nonumber\\
Q_{kl}^{\rm A} &=& \sum_{m=-2}^{2}|\langle k|Y_2^{m}|l\rangle|^2\,,
\end{eqnarray}
where $Y_2^{m}$ are the Stevens operator equavilents of the spherical harmonics for $L=2$.

\begin{figure}
\centering
\includegraphics[totalheight=7cm,angle=0]{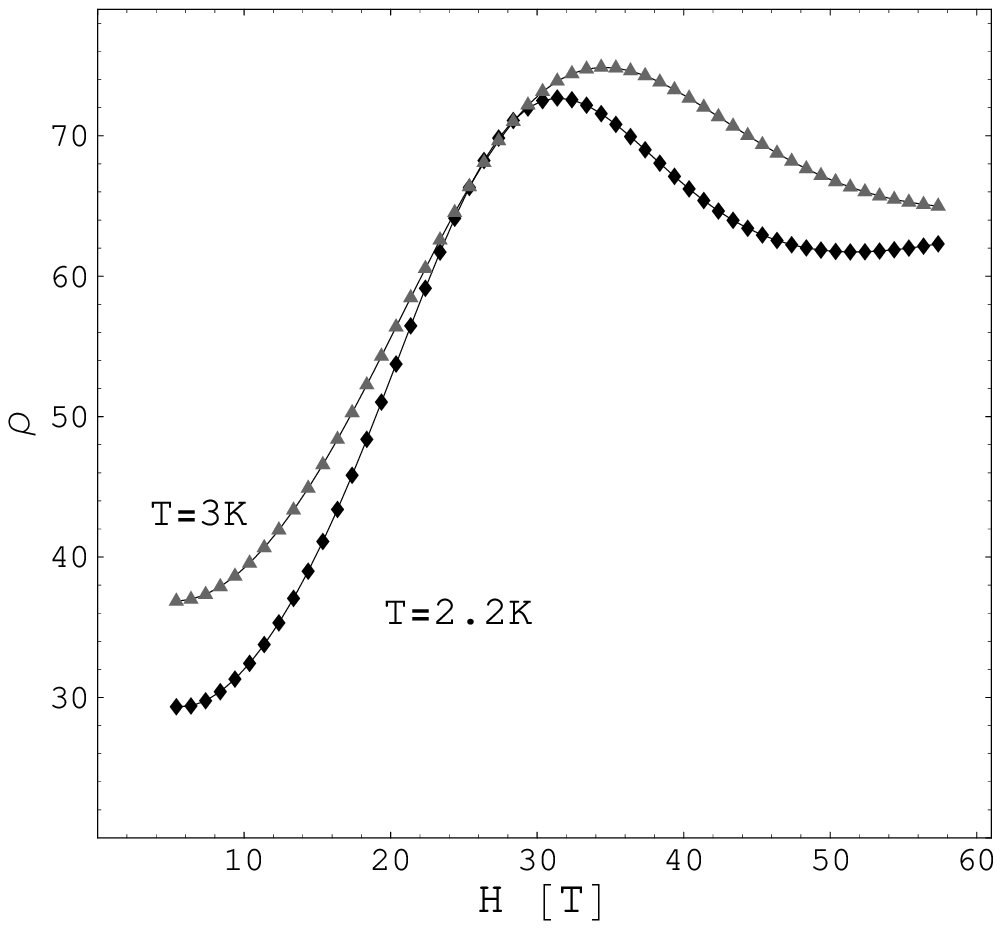}
\includegraphics[totalheight=7cm,angle=0]{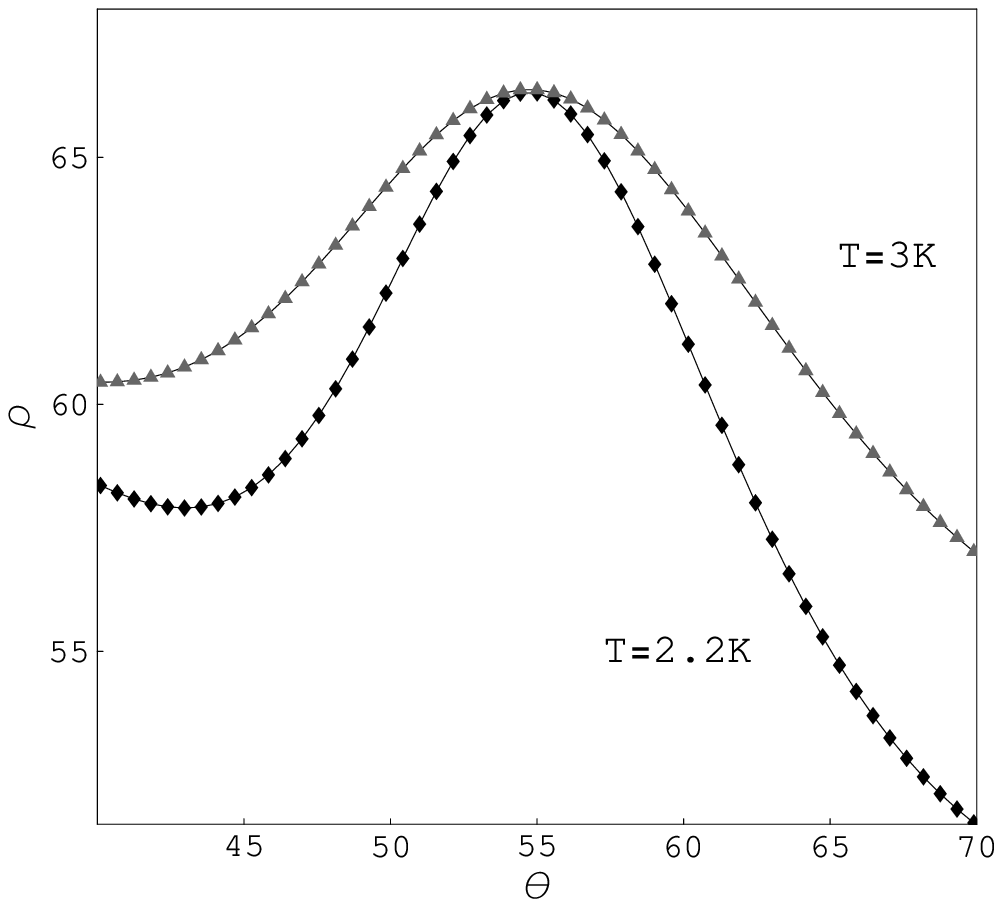}
\caption{{\sl Left:} Electrical resistivity calculated as a function of ${\bf H}\|(111)$ at $T=2.3$K and $3$K using the crystal field scheme shown in Fig.~\ref{fig:3}. {\sl Right:} Electrical resistivity calculated as a function of angle $\theta$ (see text) near the $(111)$ direction at $H=25.3$T for $T=2.2$K and $3$K. In both figures the resistivity is in arbitrary unit and $r=1/2$.}\label{fig:7}
\end{figure}

With the crystal field parameters and the corresponding level scheme presented in Fig.~\ref{fig:3}, we calculate the electrical resistivity considering the lowest six levels as a function of ${\bf H}\|(111)$. 
We used the dipole interaction parameter $\lambda_{dip}=-102.5$mK of Fig.~\ref{fig:2} for the determination of the external field ($H$) 
from the molecular field.
The octupolar interaction does not contribute to the molecular field in our model of eq.(\ref{eq:int3}).
The result is shown in the left part of Fig.~\ref{fig:7}. Due to the level crossing at $H\sim 23$T, the resistivity shows a corresponding peak 
at low temperatures. 
This peak can be interpreted as 
an enhanced contribution to 
$P_{kl}$ from degenerate states\cite{maple}. 
Let $|k\rangle$ and $|l\rangle$  be the ground and the first excited states, respectively.  If higher levels can be neglected,
we obtain
\begin{eqnarray}
P_{kl}\approx \frac{\beta (E_k-E_l)}{2\sinh \beta (E_k- E_l)}. \label{eq:pij}
\end{eqnarray}
This function shows a peak around $E_k-E_l=0$, which means 
an increase in the resistivity.

We calculated also the electrical resistivity 
as a function of the polar angle $\theta$ of the magnetic field.
This $\theta$ describes the rotation of the field 
from ${\bf H}\|(001)$ to ${\bf H}\|(110)$ within the  $[1,-1,0]$ plane. 
The right panel of Fig.~\ref{fig:7} shows the $\theta$ dependence of the calculated electrical resistivity for a field 
close to the level crossing point. 
The resistivity is enhanced around the $(111)$ direction ($\theta\sim 55^o$), 
and the peak becomes sharper as temperature decreases.  
In this calculation we have taken $r=1/2$.   We have checked that the results for $H$- and $\theta$-dependences hardly change if we take different values of $r$.
%
Experimentally, an even sharper resistivity peak has been found at $T=0.36$K and $H=8$T\cite{kuramochi}.  
However, according to our model, we are 
already inside the ordered phase at this $T$ and $H$ as shown in Fig.~\ref{fig:2}.  Hence our approximation assuming 
independent impurities cannot be applied for quantitative comparison with the experiment.

\section{Quadrupole patterns at low temperatures}\label{sec:g3o}

In Section 3 we derived the instability of the paramagnetic phase 
without specifying the AFQ pattern, since the transition temperature is independent of the pattern as long as they belong to the $\Gamma_3$ order-parameter space.
However, at low-temperatures the stability of the ordered phase should depend on the AFQ pattern.  Hence
it is interesting to study the basic features of the $\Gamma_3$ quadrupoles in the ordered phase.
For simplicity, we examine the problem in the ground state and take $H=0$.

First, let us consider a two-site problem with sites $A$ and $B$ 
with the interaction Hamiltonian within the $\Gamma_4^{(1)}$ subspace:
\begin{eqnarray}
{\cal H}(\Gamma_3)= \lambda_{\Gamma_3} ({\cal O}_{2,A}^{0}{\cal O}_{2,B}^{0}+{\cal O}_{2,A}^{2}{\cal O}_{2,B}^{2}). \label{eq:tsham1}
\end{eqnarray}
We take the nine-fold basis $| k\rangle_A | l\rangle_B$ for the two sites, where $k,l$ can be $0$, $+$ or $-$.
By diagonalization of ${\cal H}(\Gamma_3)$
we find
that the nine-fold degeneracy splits into a six- and a three-fold multiplets. With antiferro-type interaction ($\lambda_{\Gamma_3}>0$), the ground state is the six-fold multiplet with states
\begin{eqnarray}
&&|1\rangle = |0\rangle_A |+\rangle_B, \ \ 
|2\rangle = |0\rangle_A |-\rangle_B, \nonumber\\
&&|3\rangle = |+\rangle_A |0\rangle_B, \ \ 
|4\rangle = |-\rangle_A |0\rangle_B, \nonumber\\
&&|5\rangle =\frac{1}{\sqrt{2}}( |+\rangle_A |-\rangle_B-|-\rangle_A |+\rangle_B), \ \ 
|6\rangle =\frac{1}{\sqrt{2}} ( |+\rangle_A |+\rangle_B-|-\rangle_A |-\rangle_B)\label{eq:tssstates}.
\end{eqnarray}
To obtain insight into the nature of degenerate wave functions,  we associate the states of the triplet with $l=2$ spherical harmonics as
\begin{eqnarray}
|0\rangle \sim  Y_2^0 \sim 3z^2-r^2, \ \
 |+\rangle \sim  Y_2^{2} \sim (x+iy)^2, \ \
 |-\rangle \sim  Y_2^{-2} \sim (x-iy)^2.\label{eq:sphh}
\end{eqnarray} 
%
If we choose the state $|0\rangle_A$, namely $(3z^2-r^2)_A$, 
the degeneracy of $|1\rangle$, $|2\rangle$ 
allows combinations 
$|1\rangle \pm |2\rangle$ to make 
$(3z^2-r^2)_A(x^2-y^2)_B$ and $(3z^2-r^2)_A(xy)_B$ as the degenerate ground states.
Similarly, combinations $|5\rangle \pm |6\rangle$ give the other ground states as $(xy)_A(x^2-y^2)_B$ and  $(x^2-y^2)_A(xy)_B$. 
We remark that the degeneracy 6 comes from the number of ways for choice of different orbitals at two sites.  

The states $|0\rangle$, $(|+\rangle+|-\rangle)/\sqrt{2}$ and $(|+\rangle-|-\rangle)/\sqrt{2}$ are simultaneous eigenstates of
${\cal O}_2^2$ and ${\cal O}_2^0$.
The degeneracy 
of the same nature can also be found in the mean-field theory.
In fact, the two-site mean-field theory at zero temperature becomes exact because of the absence of off-diagonal elements in the Hamiltonian with respect to the basis given by eq.(\ref{eq:tssstates}).  
This situation remains the same in the bcc lattice, which can be separated into $A$ and $B$ sublattices.
Hence with only the $\Gamma_3$ intersite interaction, the ground state of the bcc lattice is degenerate.  
Recognizing this situation, it is instructive to 
consider the mean field of ${\cal H}(\Gamma_3)$ at site $B$:
\begin{eqnarray}
{\cal H}_{B}=\lambda_{\Gamma_3} \left ( \langle {\cal O}_{2}^{0}\rangle_{A} {\cal O}_{2}^{0}+\langle {\cal O}_{2}^{2}\rangle_{A} {\cal O}_{2}^{2}\right ) =\lambda_{\Gamma_3} \left ( Q_{A} {\cal O}_{2}^{0}+q_{A} {\cal O}_{2}^{2}\right ) ,\label{eq:tssham2}
\end{eqnarray}
where we have used the notations $\langle {\cal O}_{2}^{0}\rangle_A \equiv Q_A$ and $\langle {\cal O}_{2}^{2}\rangle_A \equiv q_A$.
If we choose the state $|0\rangle_A$, 
${\cal H}_{B}$ is diagonal with respect to the basis set
$\{
|\alpha_1\rangle, |\alpha_2\rangle, |\alpha_3\rangle\}=\{
|0\rangle,|+\rangle,|-\rangle\}$.
Namely we obtain 
the matrix representing the mean-field at $B$ site: 
\begin{eqnarray}
h_{B0}\equiv \frac{1}{\lambda_{\Gamma_3}}\{
\langle \alpha_k|{\cal H}_B| \alpha_l\rangle
\}
&=& 
2 c   \left( \begin{array}{ccc}
                2  & 0 & 0\\
          0  & -1 & 0\\
          0  & 0 &  -1
     \end{array}
       \right), \label{eq:matr1}
\end{eqnarray}
where $c= \left(14/\sqrt{3}\right)^2  (a_Q^2+b_Q^2)$ and we have used   $Q_A=28/\sqrt{3}a_Q$ and $q_A=-28/\sqrt{3}b_Q$. 
%
We note in eq.(\ref{eq:matr1}) that 
the splitting into a singlet and a doublet is due to 
the mixing of moments ${\cal O}_2^2$ and ${\cal O}_2^0$ in the molecular field. 
This quadrupolar field can be expressed as
\begin{eqnarray}
Q_{A} {\cal O}_{2}^{0}+q_{A} {\cal O}_{2}^{2}=28/\sqrt{3}(a_Q^2+b_Q^2){\cal C}_{2}^0\sim 3z^2-r^2.
\end{eqnarray}
%
The degeneracy in $h_{B0}$ with respect to the states $|+\rangle$ and $|-\rangle$ is also indicated in the two-site calculation as seen in eq.(\ref{eq:tssstates}). 
%

Another choice for the states on $A$ is $(|+\rangle\pm |-\rangle)_A$.
With this choice we get the following mean field at $B$ site:
\begin{eqnarray}
h_{B\pm}=
c \left( \begin{array}{ccc}
                -2  & 0 & 0\\
          0  & 1 & \pm 3\\
          0  & \pm 3 &  1
     \end{array}
       \right), 
\label{eq:matr2}
\end{eqnarray}
where the basis is the same as before: $\{
|0\rangle,|+\rangle,|-\rangle\}$, and we have used 
$Q_A=-14/\sqrt{3}a_Q-14b_Q$, $q_A=14/\sqrt{3}b_Q-14a_Q$. 
After diagonalization of $h_{B\pm}$, we have the same eigenvalues with those in $h_{B0}$ of eq.(\ref{eq:matr1}).
%
This means 
a degeneracy with respect to the states $|0\rangle_B$ and one of 
$(|+\rangle\mp |-\rangle)_B$.
The quadrupolar fields coming from the sublattice $A$ 
are now given by
\begin{eqnarray}
&& Q_{A} {\cal O}_{2}^{0}+q_{A} {\cal O}_{2}^{2} = -14/\sqrt{3}(a_Q^2+b_Q^2)({\cal C}_{2}^0\pm \sqrt{3}{\cal C}_{2}^2)\sim 3y^2-r^2, \ \
3x^2-r^2,
\end{eqnarray}
respectively.
Even in the cubic limit, pure staggered order of $O_2^2$ does not exist,
and the homogenous component of $O_2^0$ is always induced.
This is 
interpreted by the Landau free energy expansion in terms of order parameter components
that  the third-order term ($\Gamma_3\otimes\Gamma_3\otimes\Gamma_3$) is present where one of them is homogeneous, and two of them are staggered.

Thus, we conclude that 
a model with nearest-neighbor AFQ interaction of $\Gamma_3$ quadrupoles 
has a degeneracy with respect to different kinds of quadrupolar patterns.
This degeneracy 
comes from the situation 
that we cannot construct an AFQ order with "anti-parallel" arrangement of the quadrupolar moments within the triplet, contrary to the cases of classical $s=1/2$ spins, and AFQ pseudo spins within the $\Gamma_3$ doublet state.
In other words, we have to choose an orthogonal arrangement. 
Choosing $3z^2-r^2$-type quadrupole on one site, which can be realized by the state $|0\rangle$, there remains two possibilities for choosing an orthogonal state on the other sublattice. 
Namely, $3y^2-r^2$ and $3x^2-r^2$ are equivalent choices.
In this sense, the $\Gamma_3$ antiferro-quadrupolar ordering model within the triplet state is similar to the three-state antiferromagnetic (AFM) Potts model \cite{potts}, which possesses a macroscopic degeneracy in the ground state. 
The degeneracy in our case is broken by dipolar and/or octupolar intersite interactions.   
With applied magnetic field, on the other hand, the Zeeman energy
breaks the degeneracy even without magnetic interactions.
If a weak magnetic field is applied along the direction $(001)$,  for example, 
the ground state with the pure $\Gamma_3$ quadrupolar interaction becomes the staggered order composed of a sublattice with the state $|0\rangle$ and the other sublattice with the state $|-\rangle$.   In this case we expect a large staggered moment induced by the magnetic field.  The observed staggered field does not support this possibility \cite{iwasa10}.  Hence in actual PrFe$_4$P$_{12}$,
the degeneracy of the AFQ should have been lifted already without magnetic field.  


Experimentally, the distortion of ligands observed by X-ray or polarized neutron diffraction is $[\delta, \delta, \delta^{'}]$-type\cite{iwasa3,hao2}.
The dominance of this kind of quadrupolar moment is also indicated by the results of polarized neutron scattering\cite{hao, hao_thesis}, 
where the induced AFM moments is about 2 times larger for the field direction $(0,0,1)$ than for $(1,1,0)$. 
Taking $3z^2-r^2$-type quadrupolar field on one sublattice, 
%
the only way to keep the local $T_h$ symmetry at the other sublattice is to break the time-reversal symmetry.  Namely, 
if either $|+\rangle$ or $|-\rangle$ is realized,  experiment cannot detect the lattice distortion with an $O_2^2$ component.
However, a large magnetic moment should emerge in this case,
contrary to the experimental results.
On the other hand,  if  the time-reversal symmetry is preserved, the only possible  combinations are $(|+\rangle+|-\rangle)$ and $(|+\rangle-|-\rangle)$.  
In this case, a non-zero $x^2-y^2$-type quadrupolar field should emerge on one of the sublattices.

\section{Discussion and summary}\label{sec:sum}

In the present work we have proposed a picture  for the $\Gamma_3$ quadrupole order in 
PrFe$_4$P$_{12}$ in terms of 
the $\Gamma_1$--$\Gamma_4^{(1)}$ pseudo-quartet ground state under the $T_h$ symmetry. 
With the assumption of vanishingly small energy separation between the singlet and triplet states, the low- and the high-field phases can be obtained simultaneously. 
With the low-lying singlet-triplet CEF scheme a level crossing occurs only for the magnetic field direction $(111)$. 
This is the key point for understanding of the appearance of the high-field phase, and the resistivity enhancement occurring sharply around this field direction.
Introducing ferro-type interactions between the dipoles and $T^{\beta}$ octupoles, we have reproduced both phase boundaries qualitatively.
The magnitude of the octupolar interaction controls 
whether the high-field phase has the same (${\bf q}=(1,0,0)$) or different (${\bf q}=0$) ordering vector as that  
in the low-field phase.   
In the latter case we expect macroscopic lattice distortion in the high-field phase.
The resultant symmetry lower than trigonal should be detected by diffraction measurements.
With our 
set of ferro-type interaction parameters the Curie-Weiss temperature ranges from $\sim 4.1$ to $4.4$K, which is  
to be compared with the measured value $\Theta^{*}\approx 3.6$K.

The low-energy CEF level scheme in this compound has not been established, 
especially concerning the 
separation between the $\Gamma_1$ singlet and $\Gamma_4^{(1)}$ triplet states. 
This is related to interpretation of the inelastic neutron scattering results 
that sharp peaks appear only below the quadrupolar transition temperature. 
From our point of view the peak at $1.5$meV is ascribed to the splitting induced by the AFQ order.
Further experiments should identify the origin of the inelastic peak, especially 
by inspection of the temperature dependence near $T_Q$. 

Although the main theme of this paper has been to derive the multiple instabilities of the disorderd phase, 
we have also examined the properties of the quadrupolar order within the triplet state 
by exact two-site and mean-field calculations in the ground state. 
We have found that 
the AFQ ordering of the $\Gamma_3$ quadrupoles in the $T_h$ triplet results in a degeneracy with respect to different quadrupolar patterns. 
Other mechanism beyond the 
$\Gamma_3$ quadrupole interaction 
is necessary to resolve the degeneracy of the quadrupoles in the triplet.
At present, we cannot exclude the possibility that a tiny $x^2-y^2$-type distortion is actually realized.


In order to discuss the lattice distortion microscopically,
we have to consider the coupling of the $\Gamma_3$ quadrupoles to the lattice of Fe and/or P ions. 
Due to the bcc lattice structure, the Fe ion lattice is shared between Pr ions located on different sublattices. Because of the "non-collinear" antiferro-quadrupolar order, the distortion of the Fe ion lattice required by the quadrupolar moment on one site is not the optimal one for the other site. 

\section{Acknowledgment}

We are greatly indebted to Dr. Hiroaki Kusunose for inspiring
discussions and advices about the calculation of the phase
boundary and the generalized susceptibility. We also acknowledge discussions with Junya Otsuki about the level schemes in PrFe$_4$P$_{12}$ and PrOs$_4$Sb$_{12}$.


\begin{thebibliography}{99}
\bibitem{note2}
We neglected the contribution from the homogenous $T_{111}$ component to the effective octupolar field, because with our interaction parameters it is more than 
one order of magnitude smaller in energy than the effective dipole field coming from $J_{111}$ in magnetic field.
\end{thebibliography}
\end{document}